\documentclass[9pt,twocolumn,twoside]{osajnl}

\journal{ol} 

\setboolean{shortarticle}{true}

\title{Monolithic integration of ultraviolet microdisk lasers into photonic circuits in a III-nitride-on-silicon platform}

\author[1,2]{Farsane Tabataba-Vakili}
\author[3]{Blandine Alloing}
\author[3]{Benjamin Damilano}
\author[4]{Hassen Souissi}
\author[4]{Christelle Brimont}
\author[4]{Laetitia Doyennette}
\author[4]{Thierry Guillet}
\author[1]{Xavier Checoury}
\author[1]{Moustafa El Kurdi}
\author[3]{Sébastien Chenot}
\author[3]{Eric Frayssinet}
\author[3]{Jean-Yves Duboz}
\author[3]{Fabrice Semond}
\author[2]{Bruno Gayral}
\author[3,*]{Philippe Boucaud}

\affil[1]{Université Paris-Saclay, CNRS, C2N, 91120, Palaiseau, France.}
\affil[2]{Univ. Grenoble Alpes, CEA, IRIG-Pheliqs, 38000 Grenoble, France.}
\affil[3]{Université Côte d’Azur, CNRS, CRHEA, 06560 Valbonne, France.}
\affil[4]{L2C, Université de Montpellier, CNRS, 34095 Montpellier, France.}
\affil[*]{Corresponding author: Philippe.Boucaud@crhea.cnrs.fr}

\begin{abstract}
Ultraviolet microdisk lasers are integrated monolithically into photonic circuits using a III-nitride on silicon platform with gallium nitride (GaN) as the main waveguiding layer. The photonic circuits consist of a microdisk and a pulley waveguide terminated by out-coupling gratings. We measure quality factors up to 3500 under continuous-wave excitation. Lasing is observed from 374 nm to 399 nm under pulsed excitation, achieving low threshold energies of $0.14 ~\text{mJ/cm}^2$ per pulse (threshold peak powers of $35 ~\text{kW/cm}^2$). A large peak to background dynamic of around 200 is observed at the out-coupling grating for small gaps of 50 nm between the disk and waveguide. These devices operate at the limit of what can be achieved with GaN in terms of operation wavelength.
\end{abstract}

\setboolean{displaycopyright}{true}

\begin{document}

\maketitle

In recent years, there has been a significant interest in ultraviolet (UV) emitters, such as light emitting diodes (LEDs) and optically pumped lasers for a variety of applications including germicidal sterilization and gas sensing \cite{Nakamura1998_2, Kneissl2016}. The material system of choice for active devices in the UV spectrum is III-nitride, due to its band gap tunability from the UV-C to the visible spectrum.

III-nitride microcavity photonics is a very active field of research \cite{Butte2019}. Individual microlasers in the UV spectral range have been realized under pulsed optical pumping \cite{Selles2016_1, Selles2016_2, Zhu2020} and electrical injection \cite{Wang2019}. Photonic circuits in the UV have also recently gained popularity with potential applications including atomic clocks or precision metrology \cite{Blumenthal2020}. Several passive circuits have been demonstrated using aluminum nitride (AlN) \cite{Stegmaier2014, Soltani2016, Lu2018, Liu2018} or aluminum oxide (Al$_2$O$_3$) \cite{West2019}. A simple active circuit using aluminum gallium nitride (AlGaN) LEDs has also recently been demonstrated \cite{Floyd2019}. However, no demonstrations of active photonic circuits containing microlasers in the UV have been reported, while this would constitute an important step for the development of next generation photonic circuits. 

Meanwhile, there have been several reports of optically pumped active microlaser photonic circuits in the blue using III-nitrides \cite{TabatabaVakili2018, TabatabaVakili2019_2, To2020}, as well as efforts to combine microrings under electrical injection with photonic circuits \cite{TabatabaVakili2019_1}.

In this Letter, we report on the demonstration of active microlaser photonic circuits in the UV consisting of a microdisk, a pulley waveguide, and out-coupling gratings terminating the waveguide. 

The investigated sample was grown by metal organic chemical vapor deposition (MOCVD) on silicon (111). First, a 220 nm AlN buffer layer was grown, followed by 300 nm of GaN. Then the active region containing 5 pairs of 2 nm In$_x$Ga$_{1-x}$N/ 9 nm GaN quantum wells (QWs) (nominal indium composition $x=0.1$) was grown, followed by a 20 nm GaN cap layer. The total thickness is around 600 nm.

We fabricated microdisk photonic circuits using a process similar to the one described in Refs. \cite{TabatabaVakili2018, TabatabaVakili2019_2}. An SiO$_2$ hard mask, e-beam lithography using UV5 resist, and inductively coupled plasma (ICP) etching with CH$_2$F$_2$ and CF$_4$ gases for the SiO$_2$ and Cl$_2$ and BCl$_3$ gases for the III-nitrides were employed. The process consisted of two levels of lithography and etching. First, the microdisk, waveguide and grating couplers were defined, using proximity effect correction at the gap and a higher e-beam dose for the grating coupler. After the ICP etch, we performed a chemical treatment with AZ400K developer at $40^\circ \text{C}$ to smooth the side-walls. In the second level, we opened an area containing the waveguide to etch away the QWs in order to avoid re-absorption of the emission. Finally, the silicon was underetched using XeF$_2$ gas to provide for vertical confinement by refractive index contrast to air. We fabricated devices with $3~\mu \text{m}$ diameter disks and waveguides with an angle of $180^\circ$ around the disk with nominally 125 nm width in the proximity of the disk and 500 nm width away from the disk. The distance from the center of the microdisk to the end of the waveguide is $50~\mu \text{m}$. The grating couplers have a period of 200 nm. The gap between the disk and the waveguide is varied between 40 and 100 nm. These devices are particularly challenging to fabricate due to the fairly long suspension length of the waveguide with the $180^\circ$ bend. For larger microdisks we observe cracking of the nanotethers that hold the waveguide, which then falls. For suspension lengths smaller than $12~\mu \text{m}$ the devices are stable. Thus, our devices are at the fabrication limit of suspended photonic circuits in this platform. The reason for using a $180^\circ$ waveguide bend is the increase in coupling length that allows to have efficient coupling at larger gaps. Using finite-difference time-domain (FDTD) simulations we estimate that the critical coupling gap is 20 nm larger than for a $90^\circ$ angle (60 nm instead of 40 nm) corresponding to the maximum coupling.

\begin{figure}[htbp]
\centering
\includegraphics[width=0.9\linewidth]{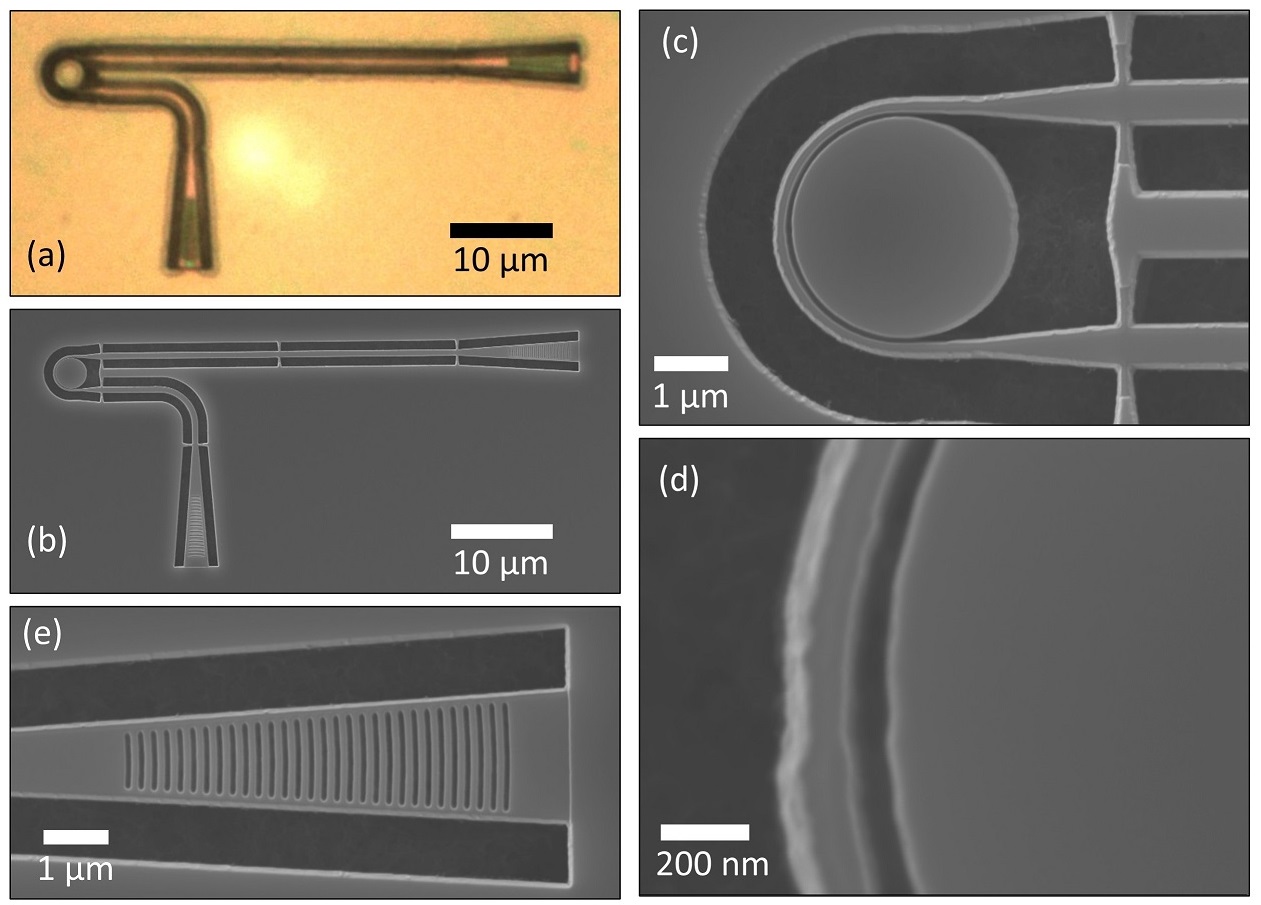}
\caption{(a) Optical microscope image and (b) SEM image of a photonic circuit. Zoom-ins of (c) the microdisk, (d) the coupling region, and (e) the grating coupler.}
\label{fig:sem}
\end{figure}

Fig. \ref{fig:sem} (a) shows an optical microscope image of a fabricated device, clearly showing the underetched areas as well as the waveguide etch through color-contrast. Fig. \ref{fig:sem} (b) shows a scanning electron microscope (SEM) image of a device, also indicating the underetch and clearly showing the nanotethers that hold the waveguide. Zoom-ins of the microdisk, the gap and the grating coupler are shown in Figs. \ref{fig:sem} (c)-(e). In Fig. \ref{fig:sem} (d) we can see the roughness of our very thin waveguides (nominally 125 nm), which is visible due to the large zoom.

\begin{figure}[htbp]
\centering
\includegraphics[width=0.8\linewidth]{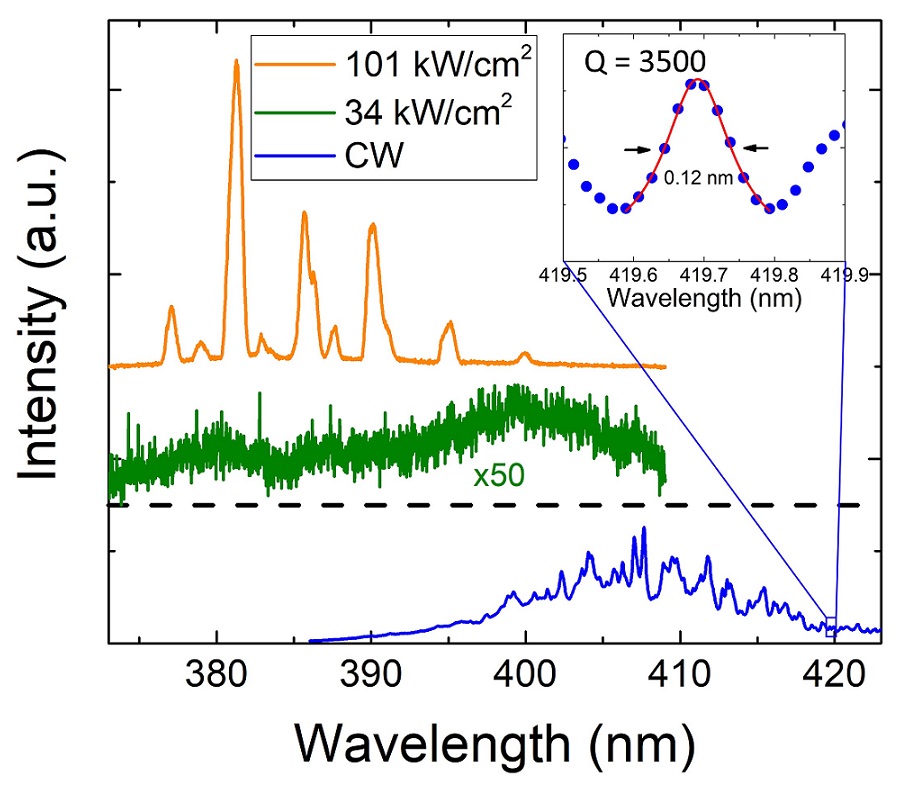}
\caption{Spectra below and above threshold taken at the out-coupling grating for devices with 100 nm gap. Bottom: Low-excitation power density CW spectrum below threshold. The inset shows the maximum Q factor of 3500 at the low energy side of the spectrum at 419.7 nm. Top: Pulsed spectra below and above threshold. The spectra are displaced along the y-axis and the CW spectrum has a different y-scale. }
\label{fig:cw}
\end{figure}

We employ a standard $\mu$-photoluminescence ($\mu$-PL) setup with a continuous-wave (CW) laser emitting at 355 nm and a 20x microscope objective that is used to pump the device and to collect the emission that is then measured by a spectrometer and a charge-coupled device (CCD). A map of the CCD allows to differentiate between the emission from the disk and from the grating by integrating over different areas. We pump the microdisks at room temperature (RT) in order to determine the loaded Q factors of our devices. The bottom part of Fig. \ref{fig:cw} shows a CW spectrum measured from the top at the out-coupling grating for a device with 100 nm gap. Many modes are visible and their identification is difficult. We observe first-order modes with an FSR of 4 nm at 407.7 nm, 411.8 nm, and 415.4 nm with $Q_{load}$ in the range of 1400 to 1800, which are limited by QW absorption. The maximum $Q_{load}$ for the 100 nm gap is 3500 at 419.7 nm (potentially the next first-order mode), which is shown in the inset in Fig. \ref{fig:cw}. We note that under low power density CW excitation the QW emission is centered at around 407 nm. The modes are not or barely visible above the microdisk under CW excitation in our top-collection setup, as whispering gallery modes (WGMs) radiate preferentially in-plane. Based on the $Q_{load}$ of the 100 nm gap device, we assume that the intrinsic Q factor $Q_{int}$ is 3500, which is in the same order of magnitude as for our previously investigated microdisks \cite{Mexis2011, Selles2016_2,TabatabaVakili2019_2}. The $Q_{load}$ at 380 nm will certainly be lower due to the GaN absorption. $Q_{int}$ might be limited by side-wall roughness or surface absorption \cite{Rousseau2018}. 

We can estimate the effective propagation losses of the microdisk using \cite{Luo2011}
\begin{equation}
    \alpha_{disk} = \frac{\lambda_0}{Q_{int}\cdot FSR \cdot r},
\end{equation}
with $\lambda_0$ the resonance wavelength, $FSR$ the free spectral range, and $r$ the radius of the disk. We obtain $\alpha=0.08~\text{dB/} \mu \text{m}$ when extrapolating at $\lambda_0 = 380~\text{nm}$, $FSR=4~\text{nm}$, $Q_{int}=3500$, and $r=1.5~\mu \text{m}$, which is two orders of magnitude larger than reported by Liu et al. for AlN microrings at 390 nm \cite{Liu2018} and in the same order of magnitude as reported by Stegmaier et al. for polycrystaline AlN waveguides at 400 nm \cite{Stegmaier2014}. When comparing with Ref. \cite{Liu2018}, we need to consider several factors. Firstly, the different waveguiding material, GaN vs. AlN: GaN will have higher losses due to larger material absorption at 380 nm, an absorption of $100~\text{cm}^{-1}$ would result in a $Q_{abs}$ of 4200. Secondly, the much higher radiation loss due to strong bending of a small microdisk ($r=1.5~\mu \text{m}$) as compared to a large microring ($r=30~\mu \text{m}$) \cite{Haus2004}. Thirdly, the different refractive index contrasts between waveguide and cladding (air vs. SiO$_2$/Al$_2$O$_3$): a smaller index contrast results in significantly lower scattering loss due to roughness. Lastly, the growth substrate, Si vs sapphire: a smaller dislocation density, which can be more easily achieved on sapphire, can induce lower internal losses. Furthermore, we employ much smaller gaps between the resonator and the waveguide than Liu et al., which renders our process more complex, but leads to better cavity to waveguide coupling.

In analogy to Ref. \cite{Liu2018}, we calculate $H_{z,s}^2/(\int H_z^2 \text{ dx dz})$, where $H_{z,s}$ is the the maximum field at the side-wall of the disk and the integral is taken over the x-z cross-section, where z is the out-of-plane direction, using FDTD. For our microdisk, we obtain $2.3 ~\mu \text{m}^{-2}$, which is an order of magnitude larger than the value obtained by Liu et al. for wide microrings \cite{Liu2018}. The higher field at the interface can be explained by the larger curvature of a small disk compared to a large ring and results in larger side-wall losses, explaining in part the lower Q factor we observe.

Lasing is observed at RT under pulsed optical conditions using a laser at 355 nm with 7 kHz repetition rate and 4 ns pulse width. Fig. \ref{fig:cw} shows pulsed spectra at peak powers of $34 ~\text{kW/cm}^2$ ($0.9 E_{th}$) and $101 ~\text{kW/cm}^2$ ($2.7 E_{th}$) measured at the out-coupling grating for a device with 100 nm gap. A strong blue-shift of around 25 nm is observed with increasing excitation power as we go from CW to pulsed excitation far above threshold. The blue-shift can be explained both by quantum confined Stark effect screening at large carrier density and by state-filling of localized states of the InGaN QWs. A smaller blue-shift was observed in our previous samples, as they were grown by molecular beam epitaxy (MBE) where less localization is observed \cite{TabatabaVakili2019_2}. Furthermore, it is very likely that excited states in the QWs are lasing. Considering the e1 (electron ground state) to hh1 (heavy hole ground state) transition at 3.05 eV or 407 nm, the center emission wavelength under low excitation CW pumping, an Indium composition of 13.5\% in the QWs is deduced from 6 band \textbf{k.p} simulations without considering carrier interactions. This transition has an oscillator strength of 0.34. The QWs are very asymmetric, which makes the cross-transitions (e2-hh1 and e1-hh2, where 2 denotes the first excited state) very intense. The oscillator strength of the e2-hh1 transition at 3.31 eV or 375 nm is 0.23. For the e1-hh2 transition at 3.13 eV or 396 nm we get an oscillator strength of 0.12. The emission thus encompasses the entire lasing spectral range of our devices.

The lasing modes are rather broad. The mode at 381.2 nm at $2.7 E_{th}$ has a full-width at half maximum (FWHM) of 1 nm. Generally under pulsed pumping, linewidths are broader than under CW pumping since the carrier density changes strongly in the microlaser during one pulse (here 4 ns). We do not observe linewidth narrowing near the threshold, since the modes are not visible below threshold due to the limited signal to noise ratio under these low duty-cycle conditions.

\begin{figure}[htbp]
\centering
\includegraphics[width=0.9\linewidth]{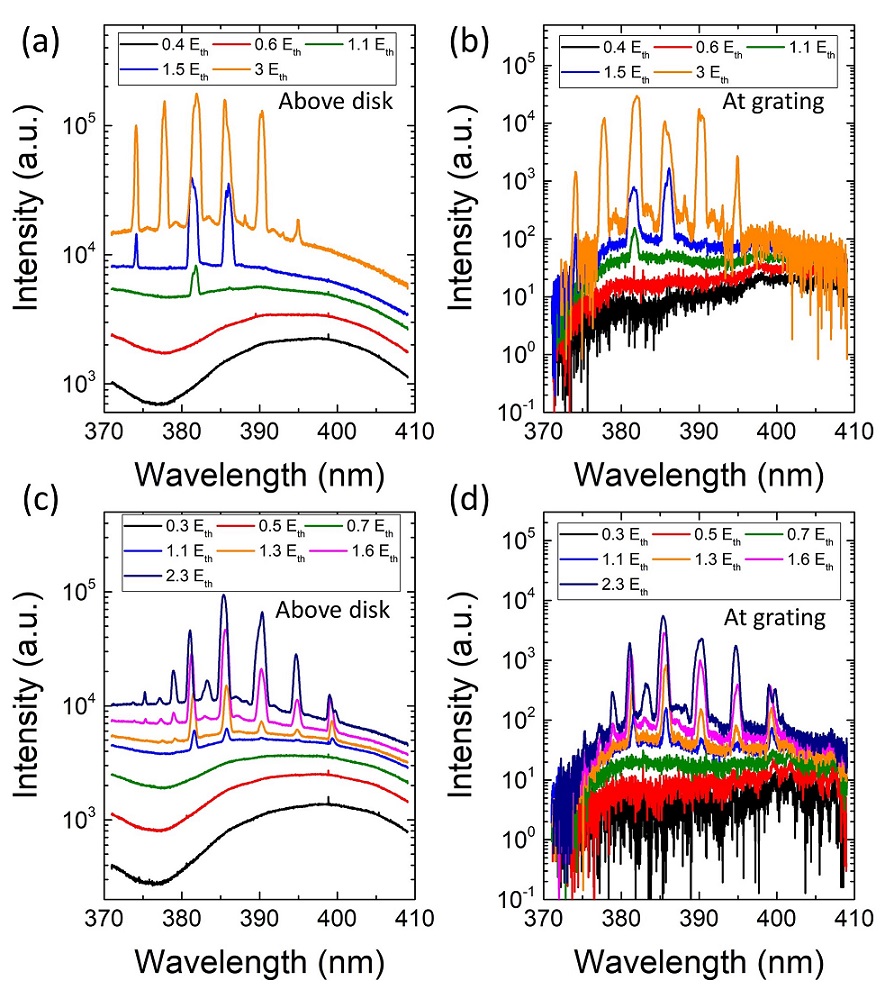}
\caption{Energy dependent pulsed optical pumping spectra of measured (a,c) above the disk and (c,d) at the grating coupler for devices with (a,b) a 50 nm gap and (c,d) an 90 nm gap. The thresholds are $0.18 ~\text{mJ/cm}^2$ per pulse for the 50 nm gap and $0.15 ~\text{mJ/cm}^2$ per pulse for the 90 nm gap.}
\label{fig:pulsed}
\end{figure}

Figs. \ref{fig:pulsed} (a) and (c) show pulse energy dependent spectra measured above the microdisk for devices with 50 nm and 90 nm gaps, respectively. Spectra measured on the same devices but at the out-coupling grating are depicted in Figs. \ref{fig:pulsed} (b) and (d). Lasing modes are observed from 374 nm to 399 nm. The lowest threshold energy of $0.14 ~\text{mJ/cm}^2$ per pulse or threshold peak power of $35 ~\text{kW/cm}^2$ is observed for the 80 nm gap (not shown). The calculated overlap of the vertical TE$_0$ mode with the QWs (excluding the barriers) is $1.0\%$. This value could be increased by using cladding layers, but those would result in a much thicker structure that makes suspended photonic circuits more difficult to fabricate. 

The first-order radial WGMs are visible for both 50 nm and 90 nm gaps (Fig. \ref{fig:pulsed}) with a spacing of 3.6 to 4.6 nm and azimuthal mode orders of $m=60$ at 374 nm to $m=54$ at 399 nm, as determined by FDTD simulations. First one and then several first order modes lase, followed by other modes of higher order, which are especially visible at high energy for the larger gaps (Figs. \ref{fig:cw} and \ref{fig:pulsed} (c,d)). We observe a very large dynamic of the peak to background emission of around 200 at $3E_{th}$ for the 50 nm gap device at the grating coupler, with only a factor 10 observed directly at the disk. The improvement of this dynamic with increased coupling is due to the fact that the WGMs radiate preferentially in-plane, while the spontaneous QW emission radiates mainly out-of-plane. In FDTD transmission simulations we obtain a maximum coupling of $70\%$ for a gap of 60 nm.

The threshold peak powers, reported here, for pulsed III-nitride microdisk lasers are lower than what we have found in literature. Simeonov et al. reported $166 ~\text{kW/cm}^2$ at 409 nm \cite{Simeonov2008} and Zhu et al. $180 ~\text{kW/cm}^2$ at 380 nm \cite{Zhu2020}. We reported lasing thresholds of $300 ~\text{kW/cm}^2$ for microdisk photonic circuits operating at 420 nm \cite{TabatabaVakili2019_2}. The threshold reduction of one order of magnitude for our samples is primarily due to a change in growth method from MBE to MOCVD for our QWs. MBE grown QWs have a lower radiative efficiency due to a higher concentration of point defects caused by the lower growth temperature \cite{Young2016}. We can compare our thresholds to pulsed electrically injected lasing reported by Wang et al. with a threshold current density of $250 ~\text{kA/cm}^2$ (or at least $800 ~\text{kW/cm}^2$ assuming operation at at least 3.2 V) at 386 nm \cite{Wang2019} for a $6.5 ~ \mu \text{m}$ thick heterostructure.

It is important to take into account that we are very close to the band gap of GaN (365 nm) in this spectral range (374 - 399 nm). Our waveguides consist to $58\%$ of GaN and to $42\%$ of AlN with $99\%$ of the TE$_0$ and $94\%$ of the TE$_1$ modes confined in the GaN layer. Consequently, there is a non-zero absorption loss that needs to be considered. The precise value of below band gap GaN absorption varies from one growth reactor to another but could be in the range of $1000 ~\text{cm}^{-1}$ at 380 nm \cite{Yu1997}. We are thus at the absolute wavelength limit of what can be achieved with a GaN waveguide. In order to decend further in wavelength, we would need to use a waveguide consisting only or mostly of AlN to reduce the absorption loss. Furthermore, using Al$_x$Ga$_{1-x}$N/Al$_y$Ga$_{1-y}$N ($y>x$) instead of InGaN/GaN  QWs would allow to reduce the emission wavelength into the UV-B and UV-C. Such photonic circuits are feasible, but remain very difficult to implement due to a need for even smaller gaps for efficient coupling as well as difficulties in growing high quality material.

\begin{figure}[htbp]
\centering
\includegraphics[width=1\linewidth]{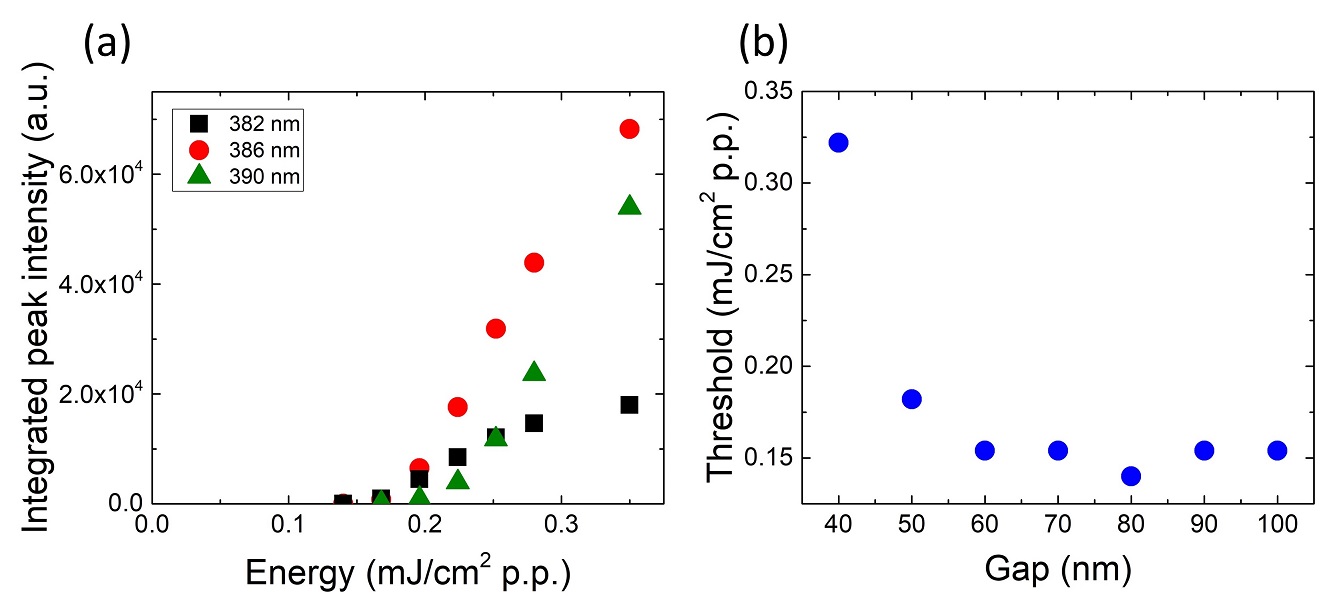}
\caption{(a) Integrated peak intensity over pulse energy for the modes at 382, 386, and 390 nm for the device with 90 nm gap shown in Fig. \ref{fig:pulsed} (c,d) and (b) Threshold energy over nominal gap size.}
\label{fig:threshold}
\end{figure}

In Fig. \ref{fig:threshold} (a) we show the integrated peak intensity over the energy per pulse for the modes at 382, 386, and 390 nm, for the device with 90 nm gap (Figs. \ref{fig:pulsed} (c,d)). The lasing thresholds at $E=0.15$ and $0.19 ~\text{mJ/cm}^2$ per pulse are clearly visible. The threshold energy as a function of nominal gap size is shown in Fig. \ref{fig:threshold} (b). We can see that the threshold decreases with increasing gap, which is to be expected, since $E_{th}\propto 1/Q_{load}$ and since $Q_{load}$ decreases with increased coupling \cite{TabatabaVakili2019_2}. For the 40 nm gap, the threshold is particularly high because the gap is not fully open, since these distances are at the limit of what can be achieved using e-beam lithography with UV5 resist.

In conclusion, we have demonstrated active microlaser photonic circuits in the UV-A spectral range using the III-nitride on silicon platform, which is very promising for both active and passive photonic components for next generation photonic integration. Our devices operate at the limit of what can be achieved using GaN as the waveguiding layer. The wavelength can be further reduced by switching from InGaN/GaN to Al$_x$Ga$_{1-x}$N/Al$_y$Ga$_{1-y}$N ($y>x$) QWs and the absorption losses can be reduced by using AlN instead of GaN/AlN for the waveguide. Possible applications of such devices could be for gas sensing \cite{Estevez2012}.


\bigskip

We thank Damir Vodenicarevic for his help with python scripts for data analysis. We also thank Sébastien Sauvage for fruitful discussions. This work was supported by the French Agence Nationale de la Recherche (ANR) under MILAGAN convention (ANR-17-CE08-0043-02). We acknowledge support by a public grant overseen by the ANR as part of the “Investissements d’Avenir” program: Labex GANEX (Grant No. ANR-11-LABX-0014). This work was also partly supported by the RENATECH network. We acknowledge the support from the technical teams at PTA-Grenoble, Nanofab (Institut Néel) and CRHEA.

\bigskip

\noindent\textbf{Disclosures.} The authors declare no conflicts of interest.

\bibliography{sample}

\bibliographyfullrefs{sample}

\end{document}